\documentclass{article}
\usepackage{graphicx}
\usepackage{amsmath}
\usepackage{amsfonts}
\usepackage{amssymb}
\usepackage{hyperref}

\begin{document}

\title{Penrose voting system and optimal quota}

\author{Wojciech S\l omczy\'{n}ski$^{a}$\\$^{a}$Institute of Mathematics, Jagiellonian University, \\
ul. Reymonta 4, 30-059 Krak\'{o}w, Poland
\and Karol \.{Z}yczkowski$^{b,c}$
\\
$^{b}$Institute of Physics, Jagiellonian University, 
\\ 
ul. Reymonta 4, 30-059 Krak\'{o}w, Poland \\
$^{c}$Center for Theoretical Physics, Polish Academy of Sciences, \\
Al. Lotnik\'{o}w 32/46, 02-668 Warszawa, Poland}

\maketitle

Abstract:

Systems of indirect voting based on the principle of qualified majority can be
analysed using the methods of game theory. In particular, this applies to the
voting system in the Council of the European Union, which was recently a
subject of a vivid political discussion. The a priori voting power of a voter
measures his potential influence over the decisions of the voting body under a
given decision rule. We investigate a system based on the law of Penrose, in
which each representative in the voting body receives the number of votes (the
voting weight) proportional to the square root of the population he or she
represents. Here we demonstrate that for a generic distribution of the
population there exists an optimal quota for which the voting power of any
state is proportional to its weight. The optimal quota is shown to decrease
with the number of voting countries.
\medskip

\medskip

PACS number(s): 02.50.Le, 89.65.-s
\medskip\newpage

Voting rules implemented by various political or economical bodies may be
studied with the help of the tools developed for many decades in game theory
\cite{Pen46,Ban65,FelMac98,FelMac04a}. We are going to analyse a special case
of indirect voting: each citizen of a given country elects a representative,
who will cast a ballot in the voting body on behalf of his electors. The
decisions of such a body are taken if certain fixed conditions characterising
qualified majority (the winning coalition) are fulfilled. For instance,
according to the agreement reached in Brussels in June 2004 and signed in Rome
in October 2004, the Council of Ministers of the European Union (EU) acting on
a proposal from the Commission or from the Union Minister for Foreign Affairs
takes its decisions if two criteria are simultaneously satisfied: a) at least
55\% of members of the Council, comprising at least fifteen of them vote
'yes', and b) these members represent Member States comprising at least 65\%
of the total population of the Union. Additionally: c) a blocking minority
must include at least four Council members, failing which the qualified
majority shall be deemed attained. The same rules apply to the European
Council when it is acting by a qualified majority (The Treaty Establishing a
Constitution for Europe, 2004, see \cite{Treaty04}).\medskip

A mathematical theory of indirect voting was initiated after World War II by
British psychiatrist and mathematician Lionel S. Penrose (1946) in the context
of a hypothetical distribution of votes in the UN General Assembly
\cite{Pen46}. He introduced the concept of a priori voting power, a quantity
measuring the ability of a participant $X$ of the voting body to influence the
decisions taken. In 1965 a similar analysis was independently carried out by
American attorney John F. Banzhaf III \cite{BalWin04b}. The voting power is
proportional to the probability that a vote cast by $X$ in a hypothetical
ballot will be decisive: a winning coalition would fail to satisfy the
qualified majority condition without $X$ or a losing coalition would start to
satisfy it with $X$. If we assume that all potential coalitions are equally
probable, then the voting power may be expressed by the
\textsl{Penrose-Banzhaf index} (\textsl{PBI}) \cite{FelMac98,FelMac04a},
called also the Banzhaf index. For convenience one often normalises the PBIs
in such a way that their sum is equal to unity. The relative voting power
should be distinguished from the voting weight: a shareholder with $51\%$ of
stocks of a company has only $51\%$ of all votes at the shareholders assembly,
but he takes $100\%$ of the voting power if the assembly votes by a simple
majority rule. Note that this approach is purely \textsl{normative}, not
descriptive: we are interested in the a priori voting power arising from the
voting procedure itself. The actual voting power depends on the polarisation
of opinion in the voting body and changes from voting to voting
\cite{GelKat02,GelKatBaf04,PajWid04,HayAkeWal06}.\medskip

To compute the PBIs of $M$ participants of a voting system which follows a
given set of rules one needs to consider all possible $2^{M}$ coalitions to
check which of them satisfies the qualified majority condition, and to count
those for which the voice of a given participant is decisive. In the case of
the EU consisting of $25$ (or in the near future $27$) states, there are more
than $33.5$ (or, respectively, $134$) millions of possible coalitions. A
game-theoretical analysis of the rules of voting in the European Council
performed along those lines shows
\cite{FelPatSil03,BalWin04a,BalWin04b,Cam04,Ple04} that the double majority
system laid down in 2003 by the European Convention attributes a much smaller
relative voting power to Spain and Poland than the earlier system accepted in
the Treaty of Nice in 2001. In this way we obtain a mathematical explanation
of the political fact that these two countries were the main opponents of the
proposed changes to the voting rules \cite{BalWin04b,Cam04,Ade06}.\medskip

To describe an algorithm of computing the PBIs assume that $\omega$ is the
number of winning coalitions, in the sense that they satisfy the qualified
majority rule adopted. There exist $2^{M-1}$ different coalitions in which a
given country can take part. Let $\omega_{x}$ denote the number of winning
coalitions that include the country $x$. Assuming that all $2^{M}$ coalitions
are equally likely we can compute the probability that a vote cast by $x$ is
decisive. This happens, if $x$ is a \textsl{critical voter} in a coalition,
i.e., the winning coalition (with $x$) ceases to fulfil the majority
requirements without $x$. The number of these cases is: $\eta_{x}=\omega
_{x}-(\omega-\omega_{x})=2\omega_{x}-\omega$. The \textsl{absolute
Penrose-Banzhaf index} is equal to the probability that $x$ is critical:
$B_{x}=\eta_{x}/2^{M-1}$. To compare these indices for decision bodies
consisting of different number of players, it is convenient to define the
\textsl{normalised Penrose-Banzhaf index}: $\beta_{x}=\left(  \sum_{x=1}%
^{M}\eta_{x}\right)  ^{-1}\eta_{x}$. Penrose mentioned in 1946 that in this
model the probability $p_{x}$ that the country $x$ is on the `winning' side
reads:%
\[
p_{x}=(\omega_{x}+\left(  2^{M-1}-(\omega-\omega_{x})\right)  )/2^{M}%
=\frac{1+B_{x}}{2}\text{ ,}%
\]
and so it is a function of the absolute Banzhaf index.\medskip

Which voting system is fairer and more accurate? A partial answer to this
question was already given by Penrose \cite{Pen46}, who deliberated principles
of an ideal representative voting system, in which every citizen of every
country has the same potential voting power. First consider direct elections
of the government (which nominates the minister voting on behalf of the entire
country in the European Council) in a state with population $N$. It is easy to
imagine that an average German citizen has smaller influence on the election
of his government than, for example, a citizen of the neighbouring Luxembourg.
Making use of the Bernoulli scheme and the Stirling approximation of the
binomials, Penrose proved that in such elections the voting power of a single
citizen decays as $1/\sqrt{N}$, given that the votes of citizens are
uncorrelated. Thus, the system of indirect voting applied to the European
Council would be representative in this sense, if the voting power of each
country behaved \textsl{proportionally to }$\sqrt{N}$, so that both factors
cancelled out. (This has a direct physical analogy with the random walk of a
diffusing particle \cite{Smo1906}.) This statement, known in the literature
under the name of the \textsl{square root law of Penrose} \cite{FelMac98}, was
independently proposed in the EU context by Laruelle and Widgr\'{e}n
\cite{LarWid98}, see \cite{LarWid96} for an earlier version. Since then
potential voting systems in the EU Council of Ministers that obey Penrose's
square root law have been analysed by many authors
\cite{BalBerGiaWid00,FelMac00,Hos00,FelMac01,LarVal02,Mob02,TiiWid02,Wid03,Ple04,Koo05,TaaHos06}%
. (Other arguments for the optimality of the square root formula can be found
in \cite{SchTor97,Mob98,BeiBovHar05,BarJac06,MaaNap06}.) Such voting
procedures has been also used in practice in other international institutions,
for example, in the Canadian Geoscience Council, the International Federation
of Operational Research Societies, the International Genetics Federation, the
International Mycological Association, and the World Federalist Movement.
However, it is not clear in general how to solve directly the \textsl{inverse
problem}, i.e., how to allocate weights and how to define qualified majority
rules to obtain required distribution of power
\cite{LarWid98,Sut00,Lee02,LinMac04,Wid04,Paj05}.\medskip

\begin{figure} [htbp]
      \begin{center} \
    \includegraphics[width=11.0cm,angle=0]{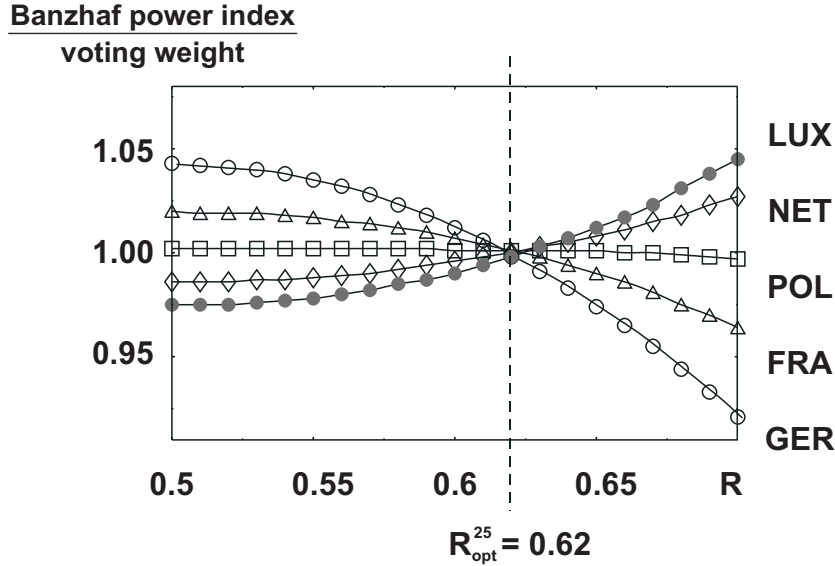}
\caption{Ratio of voting power to voting weight as a function of the quota for
five exemplary states of EU-25 (Luxembourg, the Netherlands, Poland, France,
and Germany); all functions cross near the critical point $R_{opt}^{25}=62\%$.}
\label{fig1}
\end{center}
    \end{figure}

To this end we proposed \cite{SloZyc04,ZycSloZas06} a voting system exploiting
a single criterion: the voting weight of each Member State is allocated
proportionally to the square root of its population, the decision of the
Council being taken if the sum of weights exceeds a certain \textsl{quota}
(threshold) $R$. Taking the populations $N_{x}$ ($x=1,\ldots,25$) of all 25 EU
member states as of 1 January 2003\footnote{data from \textit{EUROSTAT}: First
results of the demographic data collection for 2003 in Europe. Statistics in
focus. Population and social conditions 2004; 13; 1-7.} we analysed their
voting powers in this system as functions of the quota $R$. Fig. 1 shows the
ratio of the normalised PBIs $\beta_{x}\left(  R\right)  $ to the voting
weights proportional to $\sqrt{N_{x}}$ for five exemplary states.
Interestingly, all 25 curves (for transparency only 5 are plotted here) cross
approximately at a single point for a critical quota $R_{opt}^{25}=62\%$. Fig.
2 illustrates the dependence of the square root of the sum of square residuals
$\sigma$ between the normalised PBIs and voting weights on the value of the
threshold $R$, where%
\[
\sigma^{2}=\sum_{x=1}^{M}\left(  \beta_{x}\left(  R\right)  -\sqrt{N_{x}}%
/\sum_{y=1}^{M}\sqrt{N_{y}}\right)  ^{2}\text{ .}%
\]
Since the minimum value of this function attained for $R_{opt}^{25}$ is very
small (approximately 0.0003), we are able to work out the optimal value for the
threshold for which both the voting powers and weights coincide. For this very
choice of the quota the computed voting power of each country is practically
equal to the attributed voting weight, and so it is proportional to the square
root of the population. Hence the Penrose law is almost exactly fulfilled, and
the potential influence of every citizen of each Member State on the decisions
taken in the Council is the same. Such a voting system is not only
representative but also \textsl{transparent}: the voting powers are
proportional to the voting weights. Furthermore, the system is simple (one
criterion only), easily extendible and objective: it does not favour nor
handicap any European country. It has been christened by the media as the
`Jagiellonian Compromise'.\medskip

\begin{figure} [htbp]
      \begin{center} \
    \includegraphics[width=11.0cm,angle=0]{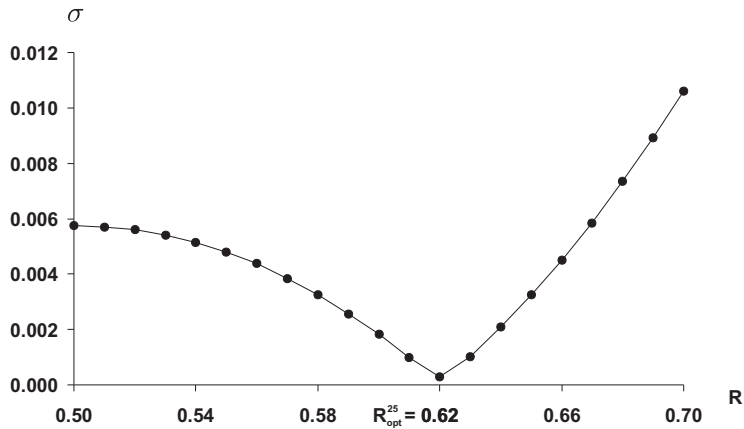}
\caption{ The cumulative residual $\sigma$ between the voting weight and power
for all EU-25 countries as dependent on the value of the threshold $R$.}
\label{fig2}
\end{center}
    \end{figure}

The main result of our work consists in the statement that the above method is
not restricted to the actual distribution of population in European countries.
Performing similar investigations for a hundred randomly chosen populations of
fictitious `Unions' containing $M$ states each, for every realisation we have
found a critical quota $R_{opt}$ at which the voting powers of all `members'
approximately coincide with the weights allocated. Thus, our method provides
in many cases a surprisingly simple solution of the inverse problem. The value
of the critical quota is realisation dependent, but for a fixed $M$ the
fluctuations are small. Moreover, the critical quota decreases with the size
of the `Union', which is rather intuitive: the more countries, the more
difficult it becomes to construct a winning coalition. For instance, for the
Council of Ministers of EU-27 (including also Roumania and Bulgaria) the
optimal quota $R_{opt}^{27}\simeq 61.4\%$, see Tab. 1.
\medskip

\begin{tabular}
[c]{|c|c|c|c|c|}\hline
Member State & Population & Voting power & Voting weight & Voting
power\\\hline
& (in millions) & (Constitution) & (Penrose) & (Penrose)\\\hline
Germany & 82.54 & 11.87 & 9.55 & 9.54\\\hline
France & 59.64 & 8.74 & 8.11 & 8.12\\\hline
United Kingdom & 59.33 & 8.69 & 8.09 & 8.10\\\hline
Italy & 57.32 & 8.44 & 7.95 & 7.96\\\hline
Spain & 41.55 & 6.37 & 6.78 & 6.79\\\hline
Poland & 38.22 & 5.89 & 6.49 & 6.50\\\hline
Roumania & 21.77 & 4.22 & 4.91 & 4.91\\\hline
Netherlands & 16.19 & 3.51 & 4.22 & 4.22\\\hline
Greece & 11.01 & 2.88 & 3.49 & 3.49\\\hline
Portugal & 10.41 & 2.80 & 3.39 & 3.39\\\hline
Belgium & 10.36 & 2.80 & 3.38 & 3.38\\\hline
Czech Republic & 10.20 & 2.78 & 3.35 & 3.35\\\hline
Hungary & 10.14 & 2.77 & 3.34 & 3.34\\\hline
Sweden & 8.94 & 2.63 & 3.14 & 3.14\\\hline
Austria & 8.08 & 2.52 & 2.98 & 2.98\\\hline
Bulgaria & 7.85 & 2.49 & 2.94 & 2.94\\\hline
Denmark & 5.38 & 2.19 & 2.44 & 2.44\\\hline
Slovakia & 5.38 & 2.19 & 2.44 & 2.44\\\hline
Finland & 5.21 & 2.17 & 2.39 & 2.39\\\hline
Ireland & 3.96 & 2.02 & 2.09 & 2.09\\\hline
Lithuania & 3.46 & 1.96 & 1.95 & 1.95\\\hline
Latvia & 2.33 & 1.82 & 1.61 & 1.61\\\hline
Slovenia & 2.00 & 1.78 & 1.48 & 1.48\\\hline
Estonia & 1.36 & 1.70 & 1.23 & 1.23\\\hline
Cyprus & 0.72 & 1.62 & 0.89 & 0.89\\\hline
Luxembourg & 0.45 & 1.59 & 0.70 & 0.70\\\hline
Malta & 0.40 & 1.58 & 0.66 & 0.66\\\hline
\end{tabular}
\medskip

Table 1. Comparison of voting power of EU-27 member states in the system of
the European Constitution and in the proposed solution (`Jagiellonian
Compromise') based on the Penrose law with the threshold $R_{opt}^{27}=61.4\%$.
\bigskip

In the limiting case as $M\rightarrow\infty$ the critical quota seems to tend
to $50\%$, consistently with the so-called Penrose limit theorem
\cite{Lin04,LinMac04}. The existence of the optimal quota was confirmed in a
recent study by Chang, Chua, and Machover \cite{ChaChuMac06} who, however,
used different measure on the set of distributions of population. Tab. 2 shows
the value of the mean critical quota as a function of the number M of members
of the voting body obtained by averaging over the random population generated
with respect to the statistical measure, i.e., the symmetric Dirichlet
distribution with Jeffreys' priors \cite{Sla99} with the density given by
\[
P\left(  x_{1,\ldots,}x_{M}\right)  =C_{M}\left(  x_{1}\cdot\ldots\cdot
x_{M}\right)  ^{-1/2}%
\]
for $x_{i}\geq0$, $\sum_{i=1}^{M}x_{i}=1$, where the normalisation constant is
expressed by the Euler gamma function, $C_{M}:=\Gamma\left(  M/2\right)
\pi^{-M/2}$. This measure on the simplex of probability distributions has been
selected since it is induced by the Fisher-Mahalanobis-Battacharyya-Rao
Riemannian metric on this set, which in turn is distinguished by being
invariant under reparametrisation \cite{Ama85}.\medskip

\begin{tabular}
[c]{|c|c|c|c|c|c|c|c|c|c|}\hline
$M$ & $10$ & $12$ & $14$ & $16$ & $18$ & $20$ & $22$ & $24$ & $26$\\\hline
$R_{opt}^{M}$ & 66.0\% & 65.8\% & 64.6\% & 64.4\% & 63.4\% & 63.1\% & 62.6\% &
62.0\% & 61.4\%\\\hline
\end{tabular}
\medskip

Table 2. Average optimal threshold $R_{opt}^{M}$ as a function of the number
of states $M$.
\medskip

The above result has a simple practical meaning: for a given number of states
$M$, choosing weights proportional to the square root of the population and
the quota in the close vicinity of $R_{opt}^{M}$ we assure that the system is
(according to the Penrose law) nearly optimally representative, since the
voting power of each country becomes proportional to the square root of its
population, and so the voting power of every citizen of each state is nearly
the same.\medskip

\begin{figure} [htbp]
      \begin{center} \
    \includegraphics[width=11.0cm,angle=0]{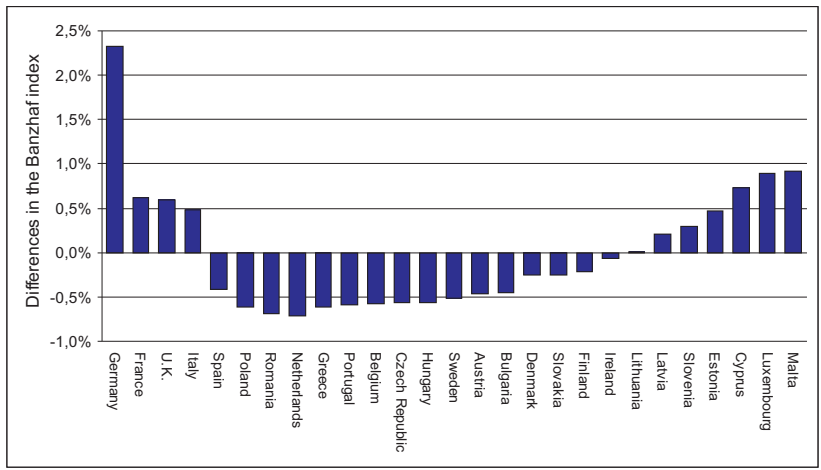}
\caption{Differences in voting power in the EU-27 Council between the system of
the European Constitution and the proposed solution (`Jagiellonian
Compromise') based on the Penrose law with $R_{opt}^{27}=61.4\%$.
The member states are ordered according to their population.} 
\label{fig3}
\end{center}
    \end{figure}

The representative voting system based on the square root law of Penrose and
the appropriate choice of optimal quota may be used as a reference point to
analyse the rules established by politicians. Fig. 3 presents a comparison of
the voting power (measured by the PBI) of EU members according to the system
accepted in Brussels in June 2004 (applied to EU-27, including also Roumania
and Bulgaria) and according to the Penrose solution with the optimal quota
$R_{opt}^{27}=61.4\%$, see \cite{Bil04,Bob04,FelMac04b,Kir04,Soz04} for
similar analyses. The double majority rule is beneficial to the largest
countries (Germany, France, the United Kingdom, and Italy), due to the `per
capita' criterion, and to the smallest countries (from Latvia to Malta), for
which the condition `per state' plays a key role. Since the largest and the
smallest countries gain relative voting power, it is easy to see that this
occurs at the expense of all the medium-sized countries (from Spain to
Ireland), which from this point of view are handicapped by the Treaty
Establishing a Constitution for Europe.

\end{document}